\documentstyle[aps,preprint,epsf]{revtex}
\def\Vec#1{\mbox{\boldmath $#1$}}
\begin{document}

\draft
\title{Hydrodynamics for inelastic Maxwell model}
\author{Hisao Hayakawa}
\address{Graduate School of Human and Environmental Studies, Kyoto
University, Kyoto 606-8501}
\date{\today}
\maketitle
\begin{abstract}
Hydrodynamic equations for an inelastic Maxwell model are derived
 from the inelastic Boltzmann equation based on a systematic
 Chapman-Enskog perturbative scheme. Transport coefficients
 appear in Navier-Stokes order have been determined as a function of the 
 restitution coefficient $e$, which cannot be defined for small $e$ as a
 result of the high energy tail of the velocity distribution function
 obeying a power law. The
 dispersion relations for the linearized equation around a homogeneous 
 cooling state have been obtained. 
\end{abstract}
\newpage
\section{Introduction}

The history of statistical mechanics begins with the gas kinetic theory of
elastically interacting particles. It is well known that
the Maxwell-Boltzmann distribution for the velocity distribution function
(VDF) plays a key role in statistical
mechanics\cite{maxwell}, and the relaxation of nonuniform dilute gases to the
equilibrium state is described by the Boltzmann equation\cite{boltzmann}.
Hydrodynamic equations  have also been derived from the Boltzmann
equation based on the Chapman-Enskog method\cite{chapman}.
On the other hand, when
there are inelastic interactions among particles, the behavior of
collections of particles is completely different from that of elastic
particles: There are no equilibrium states and any spatially homogeneous
states are
no longer achieved. 
Such a collection of the particles having inelastic interactions is
called the granular gas whose physical realization can be observed in 
rings of planets, small planets, suspended particles in fluidized beds,
and rapid granular flows {\it etc.}\cite{granular-gas}.

A proper set of hydrodynamic equations for granular fluids depend on 
situations. 
In some cases, granular particles are condensate and the fluid obeys
Bagnolds' scaling in which shear stress is proportional to the square of 
strains. This flow is called a 'frictional flow' and 
achieved in many situations under the gravity\cite{bagnolds}. 
The derivation of Bagnolds' scaling is not difficult if we adopt a 
phenomenology\cite{th-bag} but is complicated if we wish to derive it from a
microscopic motion of particles.
On the other hand, 
there are some flows dominated by binary collisions among particles. We
call this flow a 'collisional flow' or  a 'rapid granular flow' 
and its constitutive equation is
similar to that of 
the Navier-Stokes equation, {\it i.e.}, the shear stress is
proportional to the strains\cite{campbell}. For rough particles near the boundaries, 
we may need to consider the effects of spins of particles\cite{mitarai},
 but for
smooth particles we do not need such the complications. 

For collisional granular flows, we may have a standard procedure to derive
hydrodynamic equations starting from the inelastic Boltzmann equation.
The most of papers assume inelastic hard-core collisions among particles.
In fact, there are some important investigations along this line.
Jenkins and Savage\cite{jenkins83} assume that the velocity distribution function obeys
the local Maxwellian, and derive a set of hydrodynamic equations. Later, 
Lun {\it et al.}\cite{lun}, and 
Jenkins and Richman\cite{richman} 
remove the ansatz of the local Maxwellian and 
 derive hydrodynamic equations. 
Recently, Sela and Goldhirsch\cite{sela} indicate the insufficiency of the Grad
expansion by Jenkins and Richman\cite{richman}, 
and they have developed a systematic
expansion of spatial inhomogeneity and small inelasticity. In practical
sense, the method developed by 
Brey {\it et al.}\cite{brey98} and Garzo and Dufty\cite{garzo} 
is the most useful, in which 
they have extended the Chapman-Enskog method to gas particles with
inelastic hard-core collisions. Their method seems to be reasonable and 
can be used for any inelasticities. However, the method adopted there
contain some unclear points; (i) All of methods contain two types of
perturbations. One is for spatially inhomogeneity and another is for
inelasticities. Although the result by Brey {\it et al.}\cite{brey98}
 is comparable with
numerical results with any inelasticities, the logical support of their method
for inelastic particles far from the elastic limit is unclear. 
In fact, they assume that the VDF 
in a homogeneous state obeys a function
with the lowest order Sonine expansion,
which should be  only valid for nearly elastic cases. (ii)
Their analysis does not contain any information of tails in VDF
 which is believed to obey an exponential function
and cannot be described by the expansion by the Sonine polynomials. 
The effect of tails may be small but it is unclear 
how it affects hydrodynamics.  

The difficulties in the analysis of inelastic hard-core particles come
from the form of the collisional integral which is proportional to the
relative speed of colliding two particles. It is known, however, that
the collisional integral of the Maxwell molecules  which has the potential
$r^{-4}$ with the relative distance $r$ of colliding two particles 
is independent of the relative
speed\cite{maxwell,chapman,text}. 
Thus, the Maxwell model has been used for the analysis of the
Boltzmann equation as the simplest model. Quite recently, this model is
applied to kinetics of inelastic particles\cite{bobylev}. 
It is remarkable that a tail 
of the scaled velocity distribution function in a
homogeneous cooling state of the inelastic Maxwell model can be obtained 
analytically\cite{italy,ben-naim,ernst-brito,ernst-brito2}. The result of their analysis in which
 the tail of VDF obeys a power law is also interesting, because this
 result is consistent with the result of simulation in fluidized beds
 which has a long-ranged hydrodynamic interactions\cite{ichiki}, 
and leads to some
 singularities in higher order moments.

The objective of this paper is to derive a set of hydrodynamic equations 
from the inelastic Boltzmann equation systematically. For this purpose, we
adopt the inelastic Maxwell model, and the Chapman-Enskog method
developed by Brey {\it et al.}\cite{brey98}.
The organization of this paper is as follow. In the next section
, we explain the derivation of hydrodynamics from the inelastic Boltzmann
equation for the inelastic Maxwell model. The result contains the transport
coefficients such as the viscosity, the thermal conductivity, and 
the 'diffusion' coefficient as a proportional constant to the density 
gradient in the heat flux. The detailed calculation for the 
homogeneous solution and the framework of the Chapman-Enskog solution,
and explicit calculation of the transport coefficients are presented in
Appendices A, B, and C, respectively. In  Section III, we discuss the linear 
stability of the homogeneous cooling state. This requires consideration of
the explicit form of VDF at the first order and some transport coefficients
in Burnett order. Appendices D and E provide the calculations of 
these quantities. 
In Section IV, we discuss our result. In particular, we demonstrate
that the scaling solution in a cooling state is equivalent to a steady 
state with the Gaussian thermostat. So the generalization of the analysis
developed in this paper and others is relevant to discuss a steady systems 
of particles suspended by 
fluid flows. 
In Section V, we will conclude our results. 

\section{Boltzmann equation and Chapman-Enskog solution}

\subsection{Framework of the Chapman-Enskog method}

In this section, we derive the Chapman-Enskog solution of inelastic
Boltzmann equation with inelastic Maxwell particles.
Here we adopt  the Chapman-Enskog method developed by 
Brey {\it et al.}\cite{brey98}
 which is the most useful in the practical sense. 
We also restrict our interest to the case of three dimensional dilute
gases. 
Thus, the argument
is parallel to that by Brey {\it et al.}\cite{brey98}. 

Let us assume low density gases of smooth identical particles with the
mass $m$, the velocity ${\bf v}$ and the diameter $\sigma$.
Hydrodynamic variables to characterize the macroscopic behavior of  
the gas are the number density $n({\bf r},t)$, the velocity field 
${\bf u}({\bf r},t)$ and the temperature field $T({\bf r},t)$
defined by
\begin{eqnarray}
n({\bf r},t)&=& \int d{\bf v}f({\bf r},{\bf v}, t), \\
n({\bf r},t){\bf u}({\bf r},t)&=& \int d{\bf v} {\bf v}f({\bf r},{\bf v}, t) , \\
\frac{3}{2}n({\bf r},t) T({\bf r},t)&=& 
\int d{\bf v}\frac{1}{2}m {\bf V}^2 f({\bf r},{\bf v}, t),
\label{temp}
\end{eqnarray}
where ${\bf V}\equiv {\bf v}-{\bf u}$ and $f({\bf r},{\bf v}, t)$ is the
position and velocity distribution function.

The distribution function $f({\bf r},{\bf v}, t)$ 
in our system obeys the
inelastic Boltzmann equation
\begin{equation}\label{b1}
(\partial_t+{\bf v}\cdot\nabla)f=J[f,f].
\end{equation}
The collisional integral in the inelastic Maxwell model becomes
\begin{equation}\label{b2}
J[f,h]=
\sigma^2\chi\displaystyle\sqrt{\frac{T_0}{m}}\int d{\bf v}_1
\int d\Vec{\hat\sigma}
(e^{-1}b^{-1}-1)f({\bf r},{\bf v},t)h({\bf r},{\bf v}_1,t)
\end{equation}
where $T_0$ is the characteristic temperature for the potential 
among particles, ${\bf g}={\bf v}-{\bf v}_1$, 
$\Vec{\hat\sigma}$ is the unit vector along the line
connecting centers of mass of contacting particles. The operator
$b^{-1}$ is the inverse of the collisional operator $b$ which are
defined as
\begin{eqnarray}\label{b3}
b {\bf g}&=&{\bf g}-(1+e)({\bf g}\cdot \Vec{\hat\sigma})\Vec{\hat\sigma} \\
b^{-1} {\bf g}&=& {\bf g}-
\frac{1+e}{e}({\bf g}\cdot \Vec{\hat\sigma})\Vec{\hat\sigma},
\label{b5}
\end{eqnarray}
where $e$ is the coefficient of restitution which is ranged $0<e\le 1$.
Here we assume that $e$ is a constant for the simplification of our
argument, though the actual coefficient of restitution depends on the
impact velocity\cite{impact,impact2}. 
The effects of the impact velocity dependence of $e$
to macroscopic hydrodynamics can be seen in ref.\cite{brilliantov02}.  
It should be noted that the operators $b$ and $b^{-1}$ satisfy
\begin{equation}\label{b3b}
\hat b h(\hat{\bf G},{\bf g})=h(\hat{\bf G},\hat b{\bf g})
\end{equation}
for an arbitrary function $h$,
where $\hat b$ is $b$ or $b^{-1}$, and $\hat{\bf G}=({\bf v}+{\bf v}_1)/2$.
We indicate that our model is almost the same as that for inelastic
hard-core particles. The difference appears through the simplification
of the collisional integral: The term 
$\Theta({\bf g}\cdot \Vec{\hat\sigma})({\bf g}\cdot \Vec{\hat\sigma})$
for hard-core model with the Heviside function $\Theta(x)$ is replaced
by $\chi \sqrt{T_0/m}$ with a constant $\chi$. As mentioned in
introduction this simplification is justified for the potential obeying 
$r^{-4}$ with the relative distance of colliding particles for $e=1$.
It should be noted that many of inelastic Maxwell models contain the
factor $\sqrt{T/m}$ instead of $\sqrt{T_0/m}$ in
eq.(\ref{b2})\cite{bobylev,ben-naim,ernst-brito,ernst-brito2}. 
However, the choice of $\sqrt{T/m}$ leads to inconsistent results
with the elastic Maxwell model in the limit of $e\to 1$, {\it e.g.}
the viscosity becomes proportional to $\sqrt{T}$.
 Our choice ensures the consistent result with the elastic Maxwell
 model. Namely, the viscosity is proportional to $T$.
Our model may be interpreted as follows. Particles feel the long-range 
repulsive interaction by the potential obeying $r^{-4}$, and the energy 
of colliding particles loses by eq.(\ref{b3}) 
at the instance of changing the direction
of the relative velocity.

It is easily verified that the loss of kinetic energy in each collision
is 
\begin{equation}\label{bloss}
\Delta E=-\frac{1-e^2}{4}m ({\bf g}\cdot \Vec{\hat\sigma})^2.
\end{equation}
A useful identity for an arbitrary function for an arbitrary function $h$ is
\begin{equation}\label{b4}
\int d{\bf v}h({\bf v})J[f,f]
=\chi \sigma^2\displaystyle\sqrt{\frac{T_0}{m}}\int d{\bf v}_1
\int d{\bf v} f f_1 \int d\Vec{\hat \sigma}({\bf g}\cdot\Vec{\hat\sigma})
(b-1)h({\bf v}),
\end{equation}
where $f$ and $f_1$ represent $f({\bf r},{\bf v},t)$ and 
$f({\bf r},{\bf v}_1,t)$, respectively.
From this identity we directly obtain the relations
\begin{equation}\label{b6}
 \int d{\bf v} 
\left(
\begin{array}{c}
1 \\ m {\bf v} \\ \frac{1}{2}m V^2 
\end{array}\right)
 J(f,f)
=\left(\begin{array}{c}
0 \\ {\bf 0} \\ -(1-e^2)\omega[f,f]
\end{array}\right)
\end{equation}
where
\begin{equation}\label{b7}
\omega(f,f)=\frac{\pi\chi \sqrt{m T_0} \sigma^2}{6}
\int d{\bf v}_1\int d {\bf v}_2 |{\bf v}_1-{\bf v}_2|^2f({\bf
v}_1)f({\bf v}_2).
\end{equation}
The balance equations for hydrodynamic variables are
\begin{equation}\label{b8}
D_t n+ n\nabla\cdot {\bf u}=0,
\end{equation}
\begin{equation}\label{b9}
 D_t u_i+(mn)^{-1}\nabla_jP_{ij}=0.
\end{equation}
and
\begin{equation}\label{b10}
 D_t T+ \frac{2}{3n}(P_{ij}\nabla_j u_i+\nabla\cdot {\bf q})+T\zeta=0,
\end{equation}
where $D_t=\partial_t+{\bf u}\cdot\nabla$, and 
\begin{equation}\label{b11}
\zeta[f]=(1-e^2)\frac{2}{3n T}\omega[f,f].
\end{equation}
The pressure tensor $P_{ij}$ and the heat flux ${\bf q}$ are
respectively defined by
\begin{eqnarray}
P_{ij}&=&n T \delta_{ij}+ \int d{\bf V}D_{ij}({\bf V})f({\bf r},{\bf v},t)
\\
{\bf q}&=&\int d{\bf V}{\bf S}({\bf V})f({\bf r},{\bf v},t),
\end{eqnarray}
where
\begin{eqnarray}\label{b13}
D_{ij}({\bf V})&=&m(V_iV_j-\frac{1}{3}V^2\delta_{ij}), \\
{\bf S}({\bf V})&=& \left(\frac{1}{2}m{\bf V}^2-\frac{5}{2}T\right){\bf V}.
\end{eqnarray}
Once we know the solution of the inelastic Boltzmann equation, we can obtain
the complete information of hydrodynamics. It is, however, impossible 
to get the complete
solution of the Boltzmann equation which is a nonlinear and a
differential integral equation. Thus, we need a systematic 
perturbative scheme to obtain the solution.

One of the systematic methods to obtain an approximate solution
is the Chapman-Enskog method\cite{chapman}. 
This method is regarded as
a standard one in the gas kinetic theory of elastically interacting
particles. The method assumes a solution of the Boltzmann equation to
the form
\begin{equation}\label{b14} 
f({\bf r},{\bf v},t)=f[{\bf v}|n,{\bf u},T].
\end{equation}
This means that  space and time dependences appear through
hydrodynamic variables $n,{\bf u},T$. We also assume that the solution
is nearly homogeneous and exists the small spatial variations. Thus, we 
assume that the distribution function is represented by a series 
\begin{equation}\label{b15}
f=f^{(0)}+\epsilon f^{(1)}+\epsilon^2f^{(2)}+\cdots,
\end{equation}
where $\epsilon$ is a formal expansion parameter and set to be unity
after the calculation.  The expansion parameter is assumed to be
balanced with the spatial gradients. In addition, as usual in the
Chapman-Enskog method for elastically interacting particles, the time
derivative is also expanded as $\partial_t={\partial_t}^{(0)}+\epsilon
{\partial_t}^{(1)}+\cdots$. We note
${\partial_t}^{(0)}=0$ for elastic cases.

To remove the ambiguity of the distribution function, we impose
the solvability conditions as usual
\begin{eqnarray}\label{sol1}
n({\bf r},t)&=& \int d{\bf v}f^{(0)}({\bf r},{\bf v}, t), \\
n({\bf r},t){\bf u}({\bf r},t)&=& \int d{\bf v} {\bf v}f^{(0)}({\bf r},{\bf v}, t) , \\
\frac{3}{2}n({\bf r},t) T({\bf r},t)&=& 
\int d{\bf v}\frac{1}{2}m {\bf V}^2 f^{(0)}({\bf r},{\bf v}, t).
\label{sol3}
\end{eqnarray}
From these conditions, perturbative distribution functions should be
orthogonal to $f^{(0)}$, ${\bf v}f^{(0)}$ and $V^2f^{(0)}$.

\subsection{The basic solution}

The most difficult part of the Chapman-Enskog method for inelastic
particles is to obtain the zeroth order solution of the Boltzmann
equation
\begin{equation}\label{b16}
{\partial_t}^{(0)}f^{(0)}=J[f^{(0)},f^{(0)}].
\end{equation}
As mentioned in Introduction, the solution of (\ref{b16}) has been
obtained in these days.\cite{ben-naim,ernst-brito,ernst-brito2}
In this subsection, we summarize the parts of their results which will
be needed for our analysis.

Assuming the scaling form
\begin{equation}\label{b16_1}
f({\bf v},t)=n v_0(t)^{-3}
\tilde f({\bf c}), \quad {\bf c}={\bf V}/v_0(t)
\end{equation}
with $v_0(t)=\sqrt{2T/m}$,
eq.(\ref{b16}) for the inelastic Maxwell model becomes
\begin{equation}\label{ba7}
-\gamma x\Phi'(x)+\Phi(x)=\int_{\hat{\sigma}}\Phi(xe_+)\Phi(xe_-);
\quad \dot e_0=-\gamma e_0(t),
\end{equation}
where $\int_{\hat{\sigma}}=(1/4\pi)\int d\hat{\Vec{\sigma}}$ and
\begin{equation}\label{ba8} 
e_+=\hat{p}^2(\hat{\Vec{k}}\cdot \hat {\Vec{\sigma}})^2, \quad e_-=
1-z(\hat{\Vec{k}}\cdot \hat{\Vec{\sigma}})^2
\end{equation}
with $\hat{\Vec{k}}={\bf k}/k$,$\hat{p}=(1+e)/2$ and $z=1-(1-\hat{p})^2$. 
Here $\Phi(e_0(t)x)=\phi(x,t)$ with $e_0(t)=v_0(t)^2$ and
$\varphi(k,t)=\phi(k^2/4,t)$ is the Fourier transform of $f^{(0)}/n$.
The solution of (\ref{ba7}) including $\gamma$ 
can be obtained by the combination of 
the moment expansion and picking up the singularity (Appendix A). 
The result may 
be written as
\begin{equation}\label{ba8_1}
\Phi(x)=\sum_{n=0}^{[a]}(-x)^n\frac{\mu_n}{n!}-A x^a+\cdots ,
\end{equation}
where $a$ is an non-integer and $[a]$ is the largest integer less than
$a$. The moment $\mu_n$ is defined by
\begin{equation}\label{ba8_2}
\mu_n\equiv <c^{2n}>/\left(3/2\right)_n, \quad (a)_n\equiv\Gamma(a+n)/\Gamma(a)
\end{equation}
with the Gamma function $\Gamma(x)$ and $<c^{2n}>\equiv \int d{\bf
c}c^{2n}\tilde f({\bf c})$. The explicit results for $a$ and 
$\mu_n$ are presented in Appendix A.

The macroscopic equations at the zeroth order are given by
\begin{equation}\label{b17}
{\partial_t}^{(0)}n=0,\quad
{\partial_t}^{(0)}{\bf u}=0,\quad
T^{-1}{\partial_t}^{(0)} T=-\zeta^{(0)},
\end{equation}
where the cooling rate $\zeta^{(0)}$ is 
\begin{equation}\label{b18}
\zeta^{(0)}=(1-e^2)\frac{2}{3n T}\omega[f^{(0)},f^{(0)}].
\end{equation}
Substituting (\ref{b17}) into (\ref{b16}) we obtain
\begin{equation}\label{b19}
-\zeta^{(0)}T\partial_Tf^{(0)}=\frac{1}{2}\zeta^{(0)}
\frac{\partial}{\partial {\bf V}}\cdot({\bf V}f^{(0)})
=J[f^{(0)},f^{(0)}].
\end{equation}
The second expression in (\ref{b19}) is based on the assumption that 
$f^{(0)}$ is only a function of the velocity through 
the scaled velocity ${\bf c}$.

Here we should stress that $f^{(0)}$ can be evaluated in the inelastic
Maxwell model. The solution is determined from eq.(\ref{ba7}). It should 
be noted that the scaling function $\tilde f({\bf c})$ defined in (\ref{a3})
has a tail obeying $c^{-2a-3}$, where the tail is determined in 
eq.(\ref{a9.01}). 

\subsection{The determination of the transport coefficients by 
the Chapman-Enskog method}

The solution $f^{(0)}$ is isotropic so that the zeroth order pressure and
the heat flux are given by
\begin{equation}\label{b21}
{P_{ij}}^{(0)}=p\delta_{ij}, \quad {\bf q}^{(0)}={\bf 0}
\end{equation}
where $p=n T$ is the hydrostatic pressure.

The first order equation of the Boltzmann equation becomes
\begin{equation}\label{b22}
({\partial_t}^{(0)}+L)f^{(1)}=-({\partial_t}^{(1)}+{\bf v}\cdot\nabla)f^{(0)}
=-({D_t}^{(1)}+{\bf V}\cdot\nabla)f^{(0)}
\end{equation}
with ${D_t}^{(1)}={\partial_t}^{(1)}+{\bf u}\cdot\nabla$. Here 
the linear operator $L$ in (\ref{b22}) is defined by
\begin{equation}\label{b23}
Lf^{(1)}=-J[f^{(0)},f^{(1)}]-J[f^{(1)},f^{(0)}] .
\end{equation}
It is easy to verify that the zero eigenfunctions of (\ref{b23}) are not 
directly related to the collisional invariants, {\it i.e.}, $f^{(0)}$
and ${\bf v}f^{(0)}$ are not zero eigenfunctions, but they are zero
eigenfunctions of $L^{\dagger}$. This causes
significant differences in the perturbation method for systems of 
inelastic particles from those of
elastic 
particles\cite{chapman,text}. 
To recover the standard procedure, we need to restrict our 
interest to the case near $e=1$.\cite{sela}

Hydrodynamic equations at the first order give
\begin{equation}\label{b24}
{D_t}^{(1)}n=-n\nabla\cdot {\bf u}, 
\quad {D_t}^{(1)}{\bf u}=-(mn)^{-1}\nabla p,
\quad {D_t}^{(1)}T=-\frac{2T}{3}\nabla\cdot {\bf u}.
\end{equation}
Here we have used $\zeta^{(1)}=0$ from the symmetry consideration
of variables as in the case of the hard-core model.
Therefore eq.(\ref{b22}) becomes
\begin{eqnarray}
({\partial_t}^{(0)}+L)f^{(1)}&=&
f^{(0)}(\nabla\cdot{\bf u}-{\bf V}\cdot \nabla \ln n)
+(\partial_Tf^{(0)})
\left(\frac{2T}{3}\nabla\cdot {\bf u}-{\bf V}\cdot\nabla T\right)
\nonumber \\
& &+\left(\frac{\partial}{\partial V_i} f^{(0)}\right)
[-(mn)^{-1}\nabla_i p+{\bf V}\cdot\nabla u_i],
\label{b26}
\end{eqnarray}
where ${u}_i$ is the i-th component of ${\bf u}$.
The solution of this equation is assumed to be
\begin{equation}\label{b27}
f^{(1)}=\tilde{\Vec{A}}({\bf V})\cdot\nabla \ln T+\tilde{\Vec{B}}({\bf V})
\cdot \nabla \ln n+ \tilde C_{ij}({\bf V})\nabla_i u_j .
\end{equation}
Substituting this into (\ref{b26}), we can determine the functions 
$\tilde {\Vec{A}}, \tilde{\Vec{B}}$ and $\tilde C_{ij}$. The details of
calculation is given in Appendix B. The pressure tensor and the heat flux
become
\begin{eqnarray}\label{b28}
{P_{ij}}^{(1)}&=&-\eta
(\nabla_i u_j+\nabla_j u_i-\frac{2}{3}\delta_{ij}\nabla\cdot {\bf u}) \\
{\bf q}^{(1)}&=& -\kappa \nabla T-\mu \nabla n,
\end{eqnarray}
where $\eta$ and $\kappa$ are the shear viscosity and the thermal 
conductivity, respectively. The other transport coefficient $\mu$ appears
only is granular gases.

The calculation of the transport coefficients appeared in (\ref{b28}) has
been presented in Appendices B and C. They are given by
\begin{eqnarray}
\eta^*&\equiv& \frac{\eta}{\eta_0}
=[{\nu_{\eta}}^*-\zeta^*]^{-1} \label{b31}\\
\kappa^*&\equiv& \frac{\kappa}{\kappa_0}
=\frac{2(1+c^*(e))}{3({\nu_{\eta}}^*-\zeta^*)} \label{b32}\\
\mu^* &\equiv& \frac{n \mu}{\kappa_0 T}
=\frac{1}{{\nu_{\eta}}^*-2\zeta^*}\left(\frac{\zeta^*\kappa}{\kappa_0}
+\frac{1}{3}c^*(e)\right).   \label{b33}
\end{eqnarray}
Here $\eta_0$ and $\kappa_0$ are the elastic values of the shear
viscosity and the heat conductivity, respectively. Their values are
\begin{eqnarray}
\eta_0&=& \frac{T}{3n \sigma^2 A_m}\displaystyle\sqrt{\frac{2m}{T_0}}, \\
\kappa_0 &=& \frac{15 \eta_0}{4m},
\end{eqnarray}
with the numerical coefficient $A_m\simeq 1.3700$\cite{chapman2}. 
Note that the above
expression can be obtained in terms of the exact perturbative
calculation. The collision frequency $\nu_0$ is defined by $p/\eta_0$
and its value is 
\begin{equation}\label{nu0}
\nu_0=3A_m n\sigma^2\displaystyle\sqrt{\frac{T_0}{m}}.
\end{equation}
The dimensionless functions appear in eqs.(\ref{b31})-(\ref{b33}) are 
given by
\begin{eqnarray}
\zeta^*(e)&\equiv& \frac{\zeta^{(0)}}{\nu_0}
=(1-e^2)\frac{2\omega[f^{(0)},f^{(0)}]}{3n\nu_0 T} , \label{b34} \\
c^*(e)&\equiv& \frac{8}{15}\left[(\frac{m}{2T})^2\frac{1}{n}\int d{\bf V}V^4f^{(0)}
-\frac{15}{4}\right]=2(\mu_2-1) , \label{b35} \\
{\nu_{\eta}}^*&=&\frac{\int d{\bf V}D_{ij}({\bf V})L\tilde{C}_{ij}({\bf V})}
{\nu_0\int d{\bf V}D_{ij}({\bf V})\tilde{C}_{ij}({\bf V})} ,
\label{b36} \\
{\nu_{\kappa}}^*&=&
\frac{\int d{\bf V}{\bf S}({\bf V})\cdot 
L\tilde{\Vec{A}}({\bf V})}
{\nu_0\int d{\bf V}{\bf S}({\bf V})\cdot\tilde{\Vec{A}}({\bf V})},
\label{b36b} \\
{\nu_{\mu}}^*&=&
\frac{\int d{\bf V}{\bf S}({\bf V})\cdot L\tilde{\Vec{B}}({\bf V})}
{\nu_0\int d{\bf V}{\bf S}({\bf V})\cdot \tilde{\Vec{B}}({\bf V})} .
\label{b37}
\end{eqnarray}
To obtain explicit forms of the transport coefficients we need to have 
the expressions for $\tilde{\Vec A}$, $\tilde{\Vec B}$ and $\tilde
C_{ij}$. The advantage of the inelastic Maxwell model is that we can 
calculate them exactly within this perturbation scheme.
From the solvability conditions (\ref{sol1})-(\ref{sol3}),
 the leading terms of the expression are
$\tilde{\Vec A}\propto \tilde{\Vec B}\propto f^{(0)}{\bf S}({\bf V})$
and $\tilde C_{ij}\propto f^{(0)}(V)D_{ij}({\bf V})$. Then the final 
expressions are given by
\begin{eqnarray} 
\zeta^*(e)&=& \frac{5}{12}(1-e^2) , \label{b38} \\
c^*(e)&=& \frac{12(1-e)^2}{3e^2-6e-5} , \label{b39} \\
{\nu_{\eta}}^*&=& \frac{(1+e)(4-e)}{6} , \label{b40} \\
{\nu_{\kappa}}^*&=&{\nu_{\mu}}^*
=\frac{(1+e)[7\mu_3(19-11e)-20\mu_2(7-3e)+45-5e]}{24(7\mu_3-10\mu_2+5)} ,
\label{b40.1}
\end{eqnarray}
where the explicit expressions of $\mu_2$ and $\mu_3$ are presented in
(\ref{a9-4}) and (\ref{a9-5}), respectively.
Thus, the shear viscosity is given by
\begin{equation}\label{204}
\eta^*=\frac{4}{(1+e)^2}.
\end{equation}
The expressions of $\kappa^*$ and $\mu^*$ are respectively given by
\begin{equation}\label{225}
\kappa^*=\frac{8(2\mu_2-1)}{12{\nu_{\kappa}}^*-5(1-e^2)}, 
\end{equation}
and
\begin{equation}
\mu^*=\frac{4}{6{\nu_{\mu}}^*-5(1-e^2)}
\left[\frac{5(1-e^2)(2\mu_2-1)}{12{\nu_{\mu}}^*-5(1-e^2)}
+\mu_2-1\right],
\end{equation}
where ${\nu_{\kappa}}^*$ and ${\nu_{\mu}}^*$ are given in (\ref{b40.1}).
The behavior of them are 
given by Figs.1 and 2. 
It is easy to verify $\kappa^*$ and $\mu^*$ tend to 1 and 0 as $e\to 1$, 
respectively.
Although they include $\mu_3$ which diverges at
$e=e_c\simeq 0.145123$, $\kappa^*$ does not have any singularities around the
critical $e$, though the value for $e\le e_c$ does
not have any physical meaning.

The corresponding first order distribution function is also obtainable.
(See Appendix D for the derivation). The result is summarized as
\begin{equation}\label{b42}
f^{(1)}({\bf V})=-\frac{1}{nT^3}\left[\frac{4m}{5(7\mu_3-10\mu_2+5)}
{\bf S}({\bf V})\cdot(\kappa\nabla T+\mu \nabla n)
+\frac{\eta T}{\mu_2}D_{ij}({\bf V})\nabla_i u_j
\right]f^{(0)}(V).
\end{equation}

We are also interested in the stability of uniformly cooling state.
For this purpose, we need to know the form of $\zeta^{(2)}$.
As was discussed by Brey {\it et al.}\cite{brey98}, 
the important terms for the 
linear stability analysis is only two terms :
\begin{equation}\label{b43}
\zeta^{(2)}\simeq {\zeta_L}^{(2)}=\zeta_1\nabla^2T+\zeta_2\nabla^2n.
\end{equation}
Calculation of $\zeta_1$ and $\zeta_2$ is possible, as presented in
Appendix E. The results are summarized as follows:
Let us introduce the dimensionless Burnett transport coefficients
\begin{equation}\label{b66}
{\zeta_1}^*=\frac{3p}{2\kappa_0}\zeta_1,\quad
{\zeta_2}^*=\frac{3n^2}{2\kappa_0}\zeta_2.
\end{equation}
Here ${\zeta_1}^*$ and ${\zeta_2}^*$ satisfy
\begin{equation}\label{zeta1,2}
{\zeta_1}^*={c_T}^*\delta, \quad  {\zeta_2}^*={c_n}^*\delta,
\end{equation}
where
\begin{equation}\label{b43-1}
\delta=\frac{5}{12}(1-e^2)
\left(\frac{105}{8}\mu_3+\frac{15}{4}(\frac{3}{2}+\alpha)\mu_2+
\frac{9}{4}\alpha+3\beta\right),
\end{equation}
with 
\begin{equation}\label{b43-2}
\alpha=\frac{5(3\mu_2-7\mu_3)}{2(5\mu_2-3)},
\quad \beta=-\frac{15({\mu_2}^2-7\mu_3)}{4(5\mu_2-3)},
\end{equation}
and 
\begin{equation}\label{b43-1.5}
{c_T}^*\equiv \frac{{c_T}^{(2)}}{\kappa_0/p\nu_0},\quad
{c_n}^*\equiv \frac{{c_n}^{(2)}}{\kappa_0/n^2\nu_0}.
\end{equation}
Here ${c_T}^{(2)}$ and ${c_n}^{(2)}$ are introduced in (\ref{d11}).
From 
the calculation  presented in Appendix E, 
we obtain ${c_T}^*$ and ${c_n}^*$ as
\begin{equation}\label{ct}
{c_T}^*=\frac{\frac{8\kappa^*(63\mu_4-35\mu_3)}{15(7\mu_3-10\mu_2+5)}-
\frac{16}{3}\kappa^*\mu_2}
{({\nu_{\zeta}}^*-3\zeta^*)(63\mu_4+14\alpha \mu_3+4\beta \mu_2)
-\frac{16}{3}\kappa^*\mu_2}
\end{equation}
and
\begin{equation}\label{cn}
{c_n}^*=\frac{\frac{8\mu^*(63\mu_4-35\mu_3)}{15(7\mu_3-10\mu_2+5)}-
\frac{16}{3}\mu^*\mu_2}
{({\nu_{\zeta}}^*-2\zeta^*)(63\mu_4+14\alpha \mu_3+4\beta \mu_2)
-\frac{16}{3}\mu^*\mu_2}
\end{equation}
Substituting these results and (\ref{b43-1}) 
into (\ref{zeta1,2}) we obtain the explicit
form of ${\zeta_1}^*$ and ${\zeta_2}^*$.
It should be noted that ${\nu_{\zeta}}^*$ is given by (\ref{e22}).

\section{Hydrodynamic equations and their stabilities}

The results obtained in the previous section give a closed set of 
 hydrodynamic equations for $n$, ${\bf u}$ and $T$ at Navier-Stokes
 order:
\begin{eqnarray}
& & D_t n+n\nabla\cdot{\bf u}=0 \label{b56} \\
& & D_tu_i+\frac{1}{nm}\left\{\nabla_i p-\nabla_j
[\eta(\nabla_iu_j+\nabla_ju_i-\frac{2}{3}\delta_{ij}\nabla\cdot{\bf u})]
\right\}=0 \label{b57} \\
\frac{3n}{2}D_t T&+&p\nabla\cdot {\bf u}-\nabla_iu_j
[\eta(\nabla_iu_j+\nabla_ju_i-\frac{2}{3}\delta_{ij}\nabla\cdot{\bf u})]
\nonumber \\
& & -\nabla\cdot[\kappa \nabla T+\mu\nabla n ]
= -T \zeta^{(0)}-T\zeta^{(2)}.
\label{b58}
\end{eqnarray}
Here we collect terms up to the second order and set to be $\epsilon=1$.
Since we restrict our interest to the linear stability analysis of
homogeneous cooling state, as mentioned in the previous section,
$\zeta^{(2)}$ can be replaced by ${\zeta_L}^{(2)}$ as in (\ref{b43}).

For the linear stability analysis, let $\delta y\equiv y-y_H$ be the
deviation from the value in a homogeneous state, where $y$ and
$y_H$ are a hydrodynamic variable and its homogeneous value,
respectively. A set of Fourier transformed dimensionless variables are
defined by
\begin{eqnarray} 
\delta y_{k}(\tau)&=&\int d\Vec{\xi}\exp[-i{\bf k}\cdot\Vec{\xi}]
\delta y(\Vec{\xi},\tau) , \label{b59} \\
\theta_{\bf k}(\tau)&=& \frac{\delta T_{\bf k}(\tau)}{T_H(\tau)},
\quad {\bf w}_{\bf k}=\displaystyle\sqrt{\frac{m}{T_H(\tau)}}
\delta {\bf u}_{\bf k}(\tau) ,
\quad \rho_{\bf k}(\tau)=\frac{\delta n_{\bf k}(\tau)}{n_H},
\label{b60}
\end{eqnarray}
where the subscript $H$ denotes  the quantity at the homogeneous cooling 
state. $\Vec{\xi}$ and $\tau$ are the dimensionless space and time
variables,
\begin{equation}
\tau=\frac{1}{2}\int_0^tdt'\nu_H(t')=
\frac{3}{2}A_m n_H \sigma^2\displaystyle\sqrt{\frac{T_0}{m}}t,
\quad
\Vec{\xi}=\frac{\nu_H}{2}\displaystyle\sqrt{\frac{m}{T_H}}{\bf r},
\end{equation}
with $\nu_H=3A_mn_H \sigma^2\displaystyle\sqrt{\frac{T_0}{m}}.$
In terms of these variables the linearized hydrodynamic equations are
\begin{equation}\label{b62}
\partial_{\tau}\rho_{\bf k}+i k w_{k\parallel}=0,
\end{equation}
\begin{equation}\label{b63}
(\partial_{\tau}-\zeta^*+\frac{2}{3}\eta^*k^2)w_{k\parallel}
+i k(\theta_{\bf k}+\rho_{\bf k})=0
\end{equation}
\begin{equation}\label{b64}
(\partial_{\tau}-\zeta^*+\frac{1}{2}\eta^*k^2){\bf w}_{k\perp}
=0
\end{equation}
and
\begin{equation}\label{b65}
[\partial_{\tau}+\frac{5}{4}(\kappa^*-{\zeta_1}^*)k^2]\theta_{\bf k}
+[2\zeta^*+\frac{5}{4}(\mu^*-{\zeta_2}^*)k^2]\rho_{\bf k}+\frac{2}{3}
i k {\bf w}_{k\parallel}=0.
\end{equation}
Here $w_{k\parallel}$ and ${\bf w}_{k\perp}$ denote the longitudinal and 
the transversal component of ${\bf w}_{\bf k}$ defined in eq.(\ref{b60}), respectively.

Equation (\ref{b64}) is decoupled from the rest and the solution is
given by
\begin{equation}\label{b67}
{\bf w}_{k\perp}(\tau)={\bf w}_{k\perp}(0)\exp[s_{\perp}\tau]
\end{equation}
where
\begin{equation}
s_{\perp}=\zeta^*-\frac{1}{2}\eta^* k^2.
\end{equation}
This identified the degenerated shear modes.
The remaining eigenmodes have the form $\exp[s_n\tau]$ for $n=1,2,3$, 
where $s_n$ are the solutions of 
\begin{eqnarray}
s^3&+&[\frac{2}{3}\eta^*k^2-\zeta^*+\frac{5}{4}k^2(\kappa^*-{\zeta_1}^*)]s^2 
\nonumber \\ 
&+&\frac{5}{12}k^2[4+(2\eta^*k^2-3\zeta^*)(\kappa^*-{\zeta_1}^*)]s
\nonumber \\
&-&k^2[2\zeta^*-\frac{5}{4}k^2(\kappa^*-\mu^*-{\zeta_1}^*+{\zeta_2}^*)]=0.
\label{b69}
\end{eqnarray}
The dispersion relation is summarized in Fig.3. This result indicates
that two of  real parts of $s_n$ are degenerated in all regions. In addition,
the shear mode does not have the largest eigenvalue in the unstable region.

The linear hydrodynamics discussed here contains the characteristic 
wavelength;
\begin{equation}
{k_{\perp}}^c=\displaystyle\sqrt{\frac{2\zeta^*}{\eta^*}},
\quad
{k_{\parallel}}^c=\displaystyle\sqrt{\frac{8\zeta^*}{
5(\kappa^*-\mu^*-{\zeta_1}^*+{\zeta_2}^*)}},
\end{equation}
in which the uniform spatial structure larger than $({k_{\perp}}^{c})^{-1}$ and
$({k_{\parallel}}^{c})^{-1}$ is unstable. Therefore, initial long wavelength
perturbations of homogeneous cooling states grow exponentially.
To discuss the spatial structure after the instability of the homogeneous 
state, we need a proper set of nonlinear equations. Although we do not know
what the proper nonlinear 
equation is, we may expect that the time-dependent Ginzburg-Landau
 type equation may be
a candidate to characterize a nonlinear region\cite{tdgl}.

In any case, this result suggests that the system of granular gases
does not have any entropy. So the expectation that  the Tsallis 
entropy\cite{tsallis} 
can be used in granular systems is hopeless.

\section{Discussion}

In this paper, we have derived hydrodynamic equations based on the
Chapman-Enskog method. The analysis presented here is so systematic and 
straightforward that we can discuss problems in hydrodynamics of
inelastic particles in more general situations, 
{\it e.g.} how  we can apply hydrodynamics to
granular systems.

First of all, our result indicates that a transport coefficient $\mu$
diverges at  $e_c$ as a result of divergence of $\mu_3$. 
This divergences is closely related
to the high energy tail of VDF obeying a power law in the homogeneous 
cooling state. 
Our analysis contains the information of the
distribution function completely which cannot be achieved for 
systems of inelastic hard-core particles.

Second, 
 we can consider a  driven system by adding the Fokker-Planck 
operator\cite{cercignani,ernst02b}
\begin{equation}\label{IV-1}
\frac{\partial f}{\partial t}+{\bf v}\cdot \nabla f=J[f,f]+L_{FP}f,
\end{equation}
where 
\begin{equation}\label{IV-2}
L_{FP}=\gamma_0\frac{\partial}{\partial {\bf v}}\cdot [-{\bf v}+
\frac{T_B}{m}\frac{\partial}{\partial {\bf v}}].
\end{equation}
This model is physical, because the particles feel fluid drag
$-\gamma_0 {\bf v}$ and the thermal activation by the heat bath $T_B$.
This system reaches a steady state, and 
as the scaling solution in
homogeneous cooling states is equivalent to a steady solution in the
Gaussian thermostat system 
in eqs.(\ref{IV-1}) and (\ref{IV-2}) at $T_B=0$\cite{montanero}. 
It is obvious that the system at finite $T_B$ still has similar
properties to those in cooling systems. 
Although it has been recognized 
that patterns of free cooling systems\cite{zannetti} 
are similar to those in suspension
of fluidized beds\cite{tanaka},
 no connection has been discussed systematically.
In fluidized beds, there is the long ranged hydrodynamic interactions
among particles\cite{ichiki}, which means hard-core model in the
collisional integral may not be appropriate for the system with flow.
In any case, the analysis presented here is not only limited to mathematical
interest which can be solved in terms of the exact perturbation
 but also may cause 
physical interest in  applications to granular particles in fluid flows. 
We note that the high energy tails obeying a power law 
in VDF have been reported in some other 
papers of granular systems\cite{brey96,taguchi}, 
but the model in one of them\cite{brey96}
may not be appropriate as recognized by the authors themselves 
of the paper, and 
another results\cite{taguchi} could not be reproduced by any other
groups.


Third, 
one can ask if the granular temperature
introduced in (\ref{temp}) is an actual hydrodynamic variable. 
For $e=1$, $m V^2f^{(0)}\propto c^2\tilde f(c)$
 is the zero eigenfunction of $L$. 
As a result,
$T$ for a spatial homogeneous state does not have any relaxation
mechanism. In our dissipative system, $T$ is not a true hydrodynamic
variable but a quasi-hydrodynamic variable in strict sense. 
In other words we still do not have clear picture why we can assume 
the solvability condition for $T$ as in the case of $e=1$.
We, at least, need to
show that the separation of the eigenvalues of $L$ of $c^2\tilde f(c)$
from other modes. At present, though we do not have any proof of this
requirement, we may have rough picture to support this. 
Let us consider analytic functions which can be represented by polynomials
of $c$. What we need to show is the eigenvalue of $R_n(c)\tilde f(c)$
with $n$ {\it th.} order polynomial is much larger than that of 
$c^2\tilde f(c)$. This seems to be true, because $\lambda_n\propto 
\int d{\bf c}R_n(c)L[R_n(c)\tilde f(c)]\sim \mu_n$. We remember that 
there is an inequality $\mu_n\ge \mu_{n-1}$ and its equality is achieved 
at $e=1$. Since higher order moments diverge in our model. Even when $\mu_n$
is finite, $\mu_n$
with large $n$  is much larger than $\mu_2\ge\mu_1=1$.
From this simple consideration, we can expect the separation of
eigenvalues between energy and the others.
 As a result we may justify to assume that the temperature is 
a hydrodynamic variable. 
  
If we believe that non-Gaussian properties or the violations of 
the detailed balance are essential to  
steady states or scaling region of cooling states  in general dissipative
systems besides granular gases, we need investigations which is not based 
on expansions around Gaussian distribution function. In this sense, 
our systematic analysis presented here
is a good example to demonstrate how
non-Gaussian nature affects macroscopic behaviors of the system.
We hope that the analysis of 
this 'solvable' model gives some insights to understand
macroscopic behaviors in general situations of 
non-Gaussian systems or locally nonequilibrium systems.

\section{Concluding Remarks}

We have derived hydrodynamic equations for the inelastic Maxwell model
based on a systematic Chapman-Enskog method.  
We have determined 
all of the transport coefficients $\eta$, $\kappa$ and $\mu$ 
 appear in Navier-Stokes order as a function of the 
 restitution coefficient $e$. 
They cannot be defined for small $e$ as a
 result of the high energy tail of the velocity distribution function
 obeying a power law. 
We also determine the
 dispersion relations for the linearized equation around the homogeneous 
 cooling states. Finally, through the analysis in
 this paper, we clarify the limitation of conventional gas kinetics
 method, {\it i.e.} the hydrodynamics corresponding to the Navier-Stokes 
 equation exists only in $e>e_c$. 
 
{\it Note added after submission}: Immediately after my submission, I
have recieved preprint whose subject is closely related to that of this
paper\cite{santos}. Santos\cite{santos} adopts the inelastic Maxwell
model with a prefactor $\sqrt{T/m}$ instead of $\sqrt{T_0/m}$. He also
discusses steady states by adding thermostats. Although one can see 
some equivalent results between mine and his, differences exist in
(i) the model is different, and (ii) linear stability of a homogeneous
cooling state is discussed in this paper.

\vspace*{0.5cm}

The author would like to thank N. Mitarai for fruitful discussion. 
He also appreciate useful comments by S. Sasa and H-D. Kim.
He expresses his sincere gratitude to Prof. A. Santos who let him know
ref.\cite{santos}.
This work is partially supported by the Hosokawa Powder Technology Foundation,
and the Inamori Foundation. This paper is dedicated to the memory of 
Daniel C. Hong who has passed away in July, 2002. The author got an idea 
of this research during his stay at Lehigh University through discussion 
with D. C. Hong.

\begin{figure}
[htbp]
\begin{center}
 \epsfysize=7.cm
 \centerline{\epsfbox{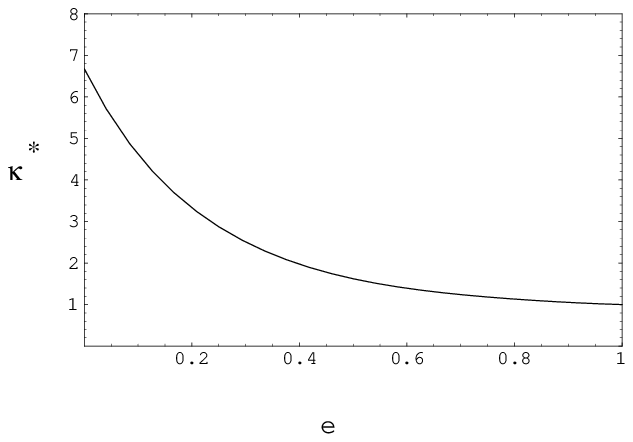}} 
\end{center}
\begin{quotation}
\caption{
$\kappa^*$ as a function of $e$.
}
\end{quotation}
\label{kappa}
%
\begin{center}
 \epsfysize=7.cm
 \centerline{\epsfbox{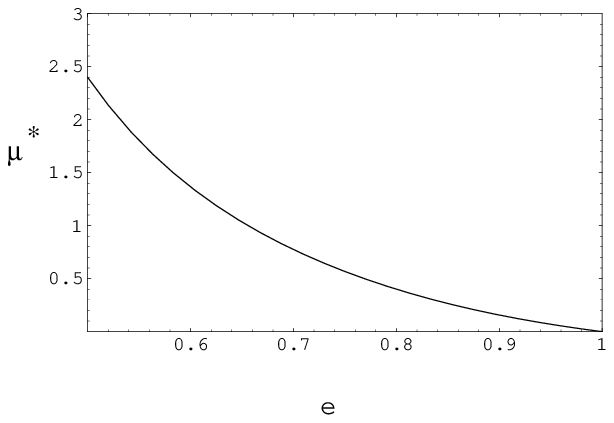}} 
\end{center}
\begin{quotation}
\caption{
$\mu^*$ as a function of $e$.
}
\end{quotation}
\label{mu}
\end{figure}

\begin{figure}[htbp]
\begin{center}
\epsfysize=10cm
 \centerline{\epsfbox{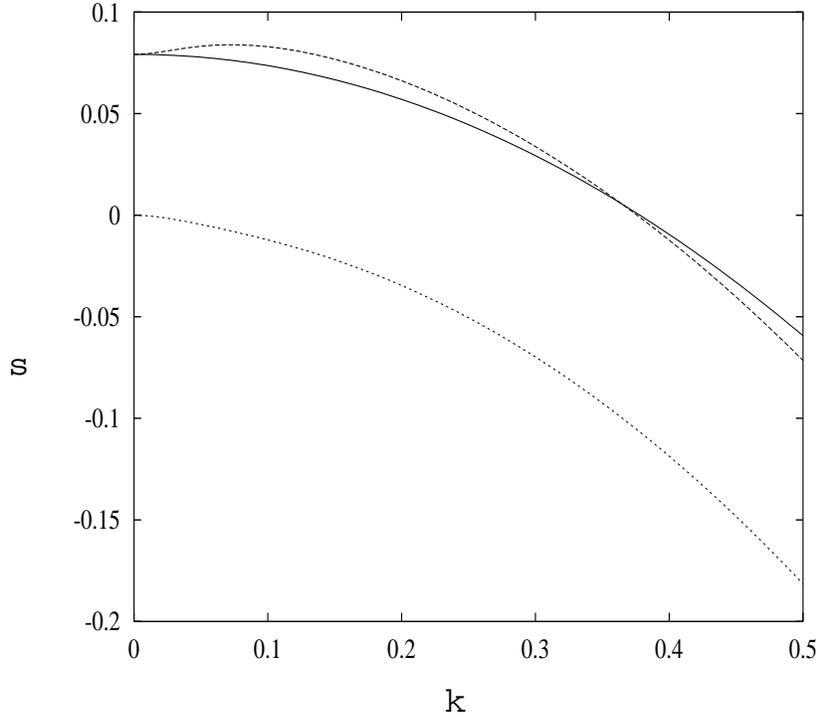}} 
\end{center}
\begin{quotation}
\caption{
The dispersion relation as a function of dimensionless $k$ at $e=0.9$.
$s(k)$ denotes the dispersion of the shear mode.
}
\end{quotation}
\label{dispersion}
\end{figure}
\appendix
\section{The determination of the basic solution}

In this Appendix we summarize the method of determination of $f^{(0)}$
obtained by some authors in these days. Now we introduce
\begin{equation}\label{a1}
f^{(0)}=n \bar f({\bf v},t), 
\quad \hat\tau=n\sigma^2\chi\displaystyle\sqrt{\frac{T_0}{m}}t.
\end{equation}
Thus, eq.(\ref{b1}) becomes
\begin{equation}\label{a2}
\partial_{\hat\tau}\bar f=I(\bar f,\bar f)\equiv
\int d\hat{\Vec{\sigma}}\int d{\bf v}_1 \frac{{\bar f}^*
{{\bar f}^*}_1}{e}-\bar f,
\end{equation}
where $\bar f^*$ and $\bar {f^*}_1$ are, respectively, the precollisional
$\bar f$ and $\bar{f}_1$, i.e. 
$\bar{f^*}=\bar f({\bf v}-\frac{1+e}{2e}({\bf g}\cdot
\hat{\Vec{\sigma}})
\hat{\Vec{\sigma}})$
and $\bar{f^*}_1=
\bar f({\bf v}_1+\frac{1+e}{2e}({\bf g}\cdot \hat{\Vec{\sigma}})
\hat{\Vec{\sigma}})$.  
Let us assume the scaling form of VDF as
\begin{equation}\label{a3}
\bar f={v_0(\tau)}^{-3}\tilde f({\bf v}/v_0(\hat\tau))
\end{equation} 
with $v_0(\hat \tau)=\sqrt{2T/m}$.
The scaling function $\tilde f({\bf c})$ satisfies the normalization
\begin{equation}\label{a4}
\int d{\bf c}\tilde f({\bf c})=1,\quad
\int d{\bf c} c^2 \tilde f({\bf c})=\frac{3}{2} .
\end{equation}
With the aid of (\ref{a4}) substituting (\ref{a3}) into (\ref{a2}) we obtain
\begin{equation}\label{a5}
\dot v_0=-\bar\gamma v_0, \quad \bar\gamma \frac{d}{d{\bf c}}\cdot
({\bf c}\tilde f({\bf c}))=I(\tilde f,\tilde f),
\end{equation}
where $\bar\gamma$ is the separation constant. Recalling (\ref{a1}) the time
evolution of $v_0(\tau)=v_0(t)$ is obtained as
\begin{equation}\label{a6}
v_0(t)=v_0(0)e^{-\bar \gamma \hat\tau}=
v_0(0)\exp\left[-\chi n\sigma^2\displaystyle\sqrt{\frac{T_0}{m}} t\right].
\end{equation}
Since the collisional integral in the inelastic Maxwell model is
independent of the relative speed, $v_0$ does not have an algebraic
decay but has the exponential decay (\ref{a6}).

Bobylev {\it et al.}\cite{bobylev} indicate
 that the Fourier transform of this model becomes 
an easy equation to discuss the behavior. Introducing the Fourier transform 
$\varphi(k,t)=\phi(\frac{1}{4}k^2,t)$ of $\bar f(v,t)$ and its scaling
form $\phi(x,t)=\Phi(e_0(t)x)$ with $e_0(t)\equiv{v_0(t)}^2$ satisfies
(\ref{ba7}).

This model has the singularity near $x=0$ of $\Phi(x)$ in eq. (\ref{ba8_1})
and the moments with $n\ge a$  
diverge. To determine $a$ we use
\begin{equation}\label{a9.01}
\gamma=\frac{2}{3}\hat p(1-\hat p)=\int_{\hat{\sigma}}[1-e_+-e_-],
\quad a \gamma=\int_{\hat{\sigma}}[1-{e_+}^a-{e_-}^a]
\end{equation}
with $\int_{\hat{\sigma}}=(1/4\pi)\int d\hat{\Vec{\sigma}}$.
From eq.(\ref{a9.01}) the exponent can be determined.
This singular term reflects on the tail of VDF obeying a power law
$\tilde f({\bf c})\sim c^{-2a-3}$.

The moment $\mu_n$ in (\ref{ba8_2}) which
is introduced in eq.(\ref{ba8_2}) satisfies
 the iterative equation 
\begin{equation}\label{a9_2}
\mu_n=\frac{1}{\gamma_n}\sum_{k=1}^{n-1}H(k,n-k)\mu_{n-k}\mu_{\bf k},
\end{equation}
where $\mu_1=1$ and ;
\begin{eqnarray}\label{a9_3}
H(m,n)&=&\left(\begin{array}{c} {m+n} \\ {m}\end{array}
\right)\beta_{2m}\hat p^{2m}F(-n,m+1/2,m+3/2;z)
\nonumber \\
&=& \hat p^{2m}\left(\begin{array}{c} {m+n} \\ {m}\end{array}
\right)\sum_{k=0}^n\left(\begin{array}{c} n \\ k \end{array}\right)
\beta_{2k+2m}(-z)^k
\end{eqnarray}
with the Gaussian Hypergeometric function $F(a,b,c;z)=
\sum_n\frac{(a)_n(b)_n}{(c)_n}\frac{z^n}{n!}$ 
and $\beta_{2m}=(1/2)_m/(3/2)_m=1/(1+2m)$.
From this expression, we can write the explicit forms of $\mu_2$,
$\mu_3$ and $\mu_4$ as
\begin{equation}\label{a9-4}
\mu_2=\frac{11-6e+3e^2}{5+6e-3e^2},
\end{equation}
\begin{equation}\label{a9-5}
\mu_3=-\mu_2\frac{115-54e+66e^2-30e^3+15e^4}{33-242e+102e^2-10e^3+5e^4},
\end{equation}
and 
\begin{equation}\label{a9-6}
\mu_4=\frac{1}{\gamma_4}[\mu_3(H(1,3)+H(3,1))+{\mu_2}^2H(2,2)]
\end{equation}
where
\begin{equation}\label{a9-7a}
H(1,3)=4\hat p^2[\frac{1}{3}-\frac{3}{5}z+\frac{3}{7}z^2-\frac{z^3}{9}],
\end{equation}
\begin{equation}\label{a9-7b}
H(3,1)=4\hat p^6[\frac{1}{7}-\frac{z}{9}],
\end{equation}
\begin{equation}\label{a9-7c}
H(2,2)=6\hat p^4(\frac{1}{5}-\frac{2}{7}z+\frac{z^2}{9})
\end{equation}
and
\begin{equation}\label{a9-7d}
\gamma_4=\frac{4}{3}z-\frac{6}{5}z^2+\frac{4z^3}{7}-\frac{z^4}{9}-
\frac{{\hat p}^8}{9}-\frac{8}{3}\hat p(1-\hat p).
\end{equation}

We note that $\mu_n$ are monotonically decreasing functions of $e$, and 
are reduced to $\mu_n=1$ at $e=1$. $\mu_2$ exists for all $e$ but
$\mu_2$ and $\mu_3 $ diverge at 0.145123.. and 0.38386.., respectively.

\section{Chapman-Enskog solution}

The Chapman-Enskog method to the Boltzmann equation is an established
method to obtain an asymptotically correct solution as a series of
spatial gradients\cite{chapman,text}. In this Appendix we explain 
the method to obtain the
Chapman-Enskog solution. It should be noted that this Appendix is parallel
to Appendix A of Brey {\it et al.}\cite{brey98} for hard-core inelastic
gases. The differences appear in the evaluation of the viscosity and
the thermal conductivity for elastic Maxwell model.

The first order equation of $f^{(1)}$ may be written as
\begin{equation}\label{aa1}
({\partial_t}^{(0)}+L)f^{(1)}={\bf A}\cdot \nabla \ln T+
{\bf B}\cdot \nabla \ln n+C_{ij}\nabla_ju_{i}.
\end{equation}
Substituting (\ref{b19}) and (\ref{b21}) into (\ref{b26}), and
comparison it with (\ref{aa1}) leads to 
\begin{equation}\label{aa2}
{\bf A}({\bf V}|n,T)=\frac{{\bf V}}{2}
\frac{\partial}{\partial {\bf V}}\cdot({\bf V}f^{(0)})
-\frac{T}{m}\frac{\partial}{\partial {\bf V}}f^{(0)},
\end{equation}
\begin{equation}\label{aa3}
{\bf B}({\bf V}|n,T)=-{\bf V}f^{(0)}-
\frac{T}{m}\frac{\partial}{\partial {\bf V}}f^{(0)}
\end{equation}
and
\begin{equation}\label{aa4}
C_{ij}({\bf V}|n,T)=\frac{\partial}{\partial V_i}(V_j f^{(0)})
-\frac{1}{3}\delta_{ij}\frac{\partial}{\partial {\bf V}}\cdot ({\bf V}f^{(0)}).
\end{equation}
The solution of eq.({\ref{aa1}}) is assumed to be the form of
(\ref{b27}). We note again that $\zeta^{(1)}=0$
directly from the symmetric considerations, since $\zeta^{(1)}$ is
a scalar. We also indicate the relation 
${\partial_t}^{(0)}\nabla\ln T=-\nabla \zeta^{(0)}=-\zeta^{(0)}\nabla \ln
n$. 

From (\ref{b27}) and (\ref{aa1}) with the help of the last equation and
(\ref{b17}) we reach
\begin{equation}\label{aa7}
\left(-\zeta^{(0)}T\partial_T+L\right)\tilde{\Vec{A}}
={\bf A}
\end{equation}
\begin{equation}\label{aa8}
\left(-\zeta^{(0)}T\partial_T+L\right)\tilde{\Vec{B}}={\bf B}+\zeta^{(0)}
\tilde{\Vec{A}}
\end{equation}
\begin{equation}\label{aa9}
\left(-\zeta^{(0)}T\partial_T+L\right)\tilde{C}_{ij}=C_{ij}.
\end{equation}
Let us calculate the viscosity at first. The pressure tensor at the
first order is written as
\begin{eqnarray}
{P_{ij}}^{(1)}&=&\int d{\bf V}D_{ij}({\bf V})\tilde{C}_{kl}\nabla_k u_l
\nonumber \\
&=&-\eta(\nabla_ju_i+\nabla_iu_j-\frac{2}{3}\delta_{ij}\nabla\cdot {\bf u}). 
\end{eqnarray}
After a long calculation we can show the relation
\begin{equation}\label{aa11}
\eta=-\frac{1}{10}\int d{\bf V}D_{ij}({\bf V})\tilde{C}_{ij}({\bf V}).
\end{equation}
Introducing
\begin{equation}\label{aa13}
\nu_{\eta}=\frac{\int d{\bf V}D_{ij}({\bf V})L\tilde{C}_{ij}({\bf V})}
{\int d{\bf V}D_{ij}({\bf V})\tilde{C}_{ij}({\bf V})}
\end{equation}
eq.(\ref{aa9}) is reduced to
\begin{equation}\label{aa14}
(-\zeta^{(0)}T\partial_T+\nu_{\eta})\eta
=-\frac{1}{10}\int d{\bf V}D_{ij}({\bf V}){C}_{ij}({\bf V}).
\end{equation}
This equation can be solved as
\begin{eqnarray}\label{aa15}
\eta&=&-\frac{1}{10(\nu_{\eta}-\zeta^{(0)})}\int
d{\bf V}D_{ij}({\bf V})\frac{\partial}{\partial V_i}(V_jf^{(0)})\nonumber\\
&=&\frac{1}{3(\nu_{\eta}-\zeta^{(0)})}\int d{\bf V}m V^2f^{(0)}
=\frac{p}{\nu_{\eta}-\zeta^{(0)}}.
\end{eqnarray}

In the elastic Maxwell model in the dilute gas, the viscosity is 
given by\cite{maxwell,chapman,chapman2}
\begin{equation}\label{max01}
\eta_0=\frac{1}{3}\displaystyle\sqrt{\frac{2m}{T_0}}\frac{T}{A_m\sigma^2}
=\frac{1}{3}\displaystyle\sqrt{\frac{2m}{T_0}}\frac{p}{n \sigma^2 A_m}
\end{equation}
and there is a relation between $\eta_0$ and the thermal conductivity
$\kappa_0$ as
\begin{equation}\label{max02}
\kappa_0=\frac{15\eta_0}{4m},
\end{equation}
where $A_m\simeq 1.3700$ is a constant, and $T_0$ is the strength of the
repulsive potential. 
 Let us introduce a
characteristic collision frequency defined by
\begin{equation}\label{aa16}
\nu_0\equiv \frac{p}{\eta_0}=3 A_m n
\sigma^2\displaystyle\sqrt{\frac{T_0}{m}}.
\end{equation}
Thus, (\ref{aa15}) becomes
\begin{equation}\label{aa17} 
\frac{\eta}{\eta_0}=\frac{1}{{\nu_{\eta}}^*-\zeta^*}
\end{equation}
where ${\nu_{\eta}}^*=\nu_{\eta}/\nu_0$ and $\zeta^*=\zeta^{(0)}/\nu_0$.

The heat flux of this order is
\begin{equation}\label{aa18}
{\bf q}^{(1)}=-\kappa \nabla T-\mu \nabla n
\end{equation}
with the transport coefficients
\begin{eqnarray}\label{aa19}
\kappa&=&-\frac{1}{3T}\int d{\bf V}{\bf S}({\bf V})\cdot
\tilde{\Vec{A}}({\bf V}), \nonumber\\
\mu &=&-\frac{1}{3n}\int d{\bf V}{\bf S}({\bf V})\cdot \tilde{\Vec{B}}({\bf V}).
\end{eqnarray}
Similar to the viscosity, $\kappa$ and $\mu$ obey
\begin{eqnarray}\label{aa20}
\kappa&=&-\frac{1}{3T(\nu_{\kappa}-\zeta^{(0)})}
\int d{\bf V}{\bf S}({\bf V})\cdot{\bf{A}}({\bf V}), \\
\mu&=&\frac{1}{\nu_{\mu}-2\zeta^{(0)}}
\left[
\zeta^{(0)}\kappa\frac{T}{n}-\frac{1}{3n}
\int d{\bf V} {\bf S}({\bf V})\cdot {\bf B}({\bf V})
\right],
\label{aa21}\end{eqnarray}
where
\begin{equation}\label{aa22}
\nu_{\kappa}=\frac{\int d{\bf V}{\bf S}({\bf V})\cdot 
L\tilde{\Vec{A}}({\bf V})}{\int d{\bf V}{\bf S}({\bf V})\cdot
\tilde{\Vec{A}}({\bf V})},\quad
\nu_{\mu}=\frac{\int d{\bf V}{\bf S}({\bf V})\cdot L\tilde{\Vec{B}}({\bf V})}
{\int d{\bf V}{\bf S}({\bf V})\cdot \tilde{\Vec{B}}({\bf V})}.
\end{equation}
Use of the formula for 
 ${\bf A}$ and ${\bf B}$ the further simplifications 
are given by
\begin{eqnarray}\label{aa23}
\frac{1}{3T}\int d{\bf V}{\bf S}({\bf V})\cdot {\bf A}({\bf V})
&=&-\frac{5nT}{2m}[1+c^*(e)]\\
\frac{1}{3n}\int d{\bf V}{\bf S}({\bf V})\cdot {\bf S}({\bf V})
&=& -\frac{5Tc^*(e)}{2m}
\label{aa24}\end{eqnarray}
where $c^*(e)$ is given by
\begin{equation}\label{b35a}
c^*(e)=
2(\mu_2-1)
\end{equation}
with the moment $\mu_2$ introduced in (\ref{a9-4}).

From these results $\kappa$ and
$\mu$ become
\begin{equation}\label{aa25}
\kappa^*=
 \frac{\kappa}{\kappa_0}=\frac{2}{3}\frac{(1+c^*(e))}{{(\nu_{\kappa}}^*-\zeta^*)},
\end{equation}
\begin{equation}\label{aa26}
\mu^*=\frac{n \mu}{T\kappa_0}=\frac{\zeta^*\kappa/\kappa_0+c^*(e)/3}
{{\nu_{\mu}}^*-2\zeta^*},
\end{equation}
where ${\nu_{\kappa}}^*=\nu_{\kappa}/\nu_0$ and ${\nu_{\mu}}^*=
\nu_{\mu}/\nu_0$.

\section{Evaluation of transport coefficients}
 
To evaluate dimensionless transport coefficients
$\zeta^*$, ${\nu_{\eta}}^*$, ${\nu_{\kappa}}^*$ and ${\nu_{\mu}}^*$ we
need to know the details form of distribution function discussed in
Appendix A. 
We should note
that we cannot separate integrals of ${\bf V}$ and ${\bf V}_1$ by 
$\hat{\bf G}$ and ${\bf g}$, since basic VDF for homogeneous states
is not the Maxwellian. 

First, we evaluate $\zeta^*$. From (\ref{b34}) $\zeta^*$ has the form
\begin{equation}\label{ab2}
\zeta^*=(1-e^2)\frac{2\omega[f^{(0)},f^{(0)}]}{3n\nu_0T}.
\end{equation}
From (\ref{b7}) and (\ref{a3}) with (\ref{a1}) we obtain
\begin{equation}\label{ab2-1}
\omega(f,f)=\frac{\pi\chi  \sigma^2n^2T \sqrt{T_0}}{3\sqrt{m}}
\int d{\bf c}_1\int d {\bf c}_2 |{\bf c}-{\bf c}_1|^2
\tilde f({\bf c})\tilde f({\bf c}_1).
\end{equation}
Here we use $\int d{\bf c}{\bf c}\tilde f({\bf c})=0$. 
Substituting this into (\ref{ab2}) with the help of (\ref{a4}) we obtain
\begin{equation}\label{ab4}
\zeta^*=(1-e^2)\frac{2\pi n \sigma^2\chi}{3\nu_0}
\displaystyle\sqrt{\frac{T_0}{m}}.
\end{equation}

The other transport coefficients
  ${\nu_{\eta}}^*$, ${\nu_{\kappa}}^*$ and ${\nu_{\mu}}^*$ 
are evaluated as follows. To lowest-order velocity dependence is
\begin{equation}\label{ab5}
\tilde{\Vec{A}}({\bf V})\propto f^{(0)}{\bf S}({\bf V}),
\quad \tilde{\Vec{B}}({\bf V})\propto f^{(0)}{\bf S}({\bf V}),
\quad \tilde{C}_{ij}({\bf V})\propto f^{(0)}D_{ij}({\bf V}).
\end{equation}
From (\ref{aa22}) and the definition of ${\nu_i}^*$ with
$i=\eta,\kappa,\mu$ we obtain
\begin{eqnarray}\label{ab6}
{\nu_{\eta}}^*&=&\frac{\int d{\bf V}D_{ij}({\bf V})L[f^{(0)}D_{ij}({\bf V})]
}{\nu_0\int d{\bf V} f^{(0)}D_{ij}({\bf V})D_{ij}({\bf V})} \nonumber \\
&=& \frac{1}{10 n T^2(1+c^*(e)/2)}
\int d{\bf V}D_{ij}({\bf V})L[f^{(0)}D_{ij}({\bf V})]
\end{eqnarray}
and 
\begin{eqnarray}\label{ab7}
{\nu_{\kappa}}^*&=&{\nu_{\mu}}^*
=\frac{\int d{\bf V}{\bf S}({\bf V})\cdot L[f^{(0)}{\bf S}({\bf V})]}
{\nu_0\int d{\bf V} f^{(0)}{\bf S}({\bf V})\cdot{\bf S}({\bf V})} \nonumber \\
&=&\frac{4m\int d{\bf V}{\bf S}({\bf V})\cdot L[f^{(0)}{\bf S}({\bf V})] }
{15nT^3\nu_0 \left(7\mu_3-10\mu_2+5\right)}.
\end{eqnarray}
Here we have used 
\begin{equation}\label{ab7-1}
\int d{\bf V} f^{(0)}{\bf S}({\bf V})\cdot{\bf S}({\bf V})
=\frac{15n T^3}{4m}\left(
 7\mu_3-10\mu_2+5\right).
\end{equation}

From the linear collision operator defined in eq.(\ref{b23}), the integral
of the form $\int YL[f^{(0)}X]$ can be transformed as
\begin{eqnarray}\label{ab9}
& & \int d{\bf v}Y({\bf v})L[X({\bf v})f^{(0)}] \nonumber \\
&=&-\sigma^2\chi\displaystyle\sqrt{\frac{T_0}{m}}\int d{\bf v}_1\int d{\bf v}
 \int d\hat{\Vec{\sigma}} Y({\bf v})
(e^{-1}b^{-1}-1)f^{(0)}({\bf v})f^{(0)}({\bf v}_1)
(X({\bf v})+X({\bf v}_1)) \nonumber \\
&=&-\sigma^2\chi\displaystyle\sqrt{\frac{T_0}{m}}\int d{\bf v}_1\int d{\bf v}
\int d\hat{\Vec{\sigma}} f^{(0)}({\bf v})f^{(0)}({\bf v}_1)X({\bf v}_1)(b-1)
(Y({\bf v})+Y({\bf v}_1)) \nonumber\\
&=& \int d{\bf v}_1X({\bf v}_1)L^{\dagger}[Y({\bf v})f^{(0)}].
\end{eqnarray}

The evaluation of the integrals of eqs.(\ref{ab6}) and (\ref{ab7}) is now
possible from the straightforward calculation.
Let us calculate ${\nu_{\eta}}^*$ at first. Noting 
$D_{ij}({\bf V})\delta_{ij}=0$, eq.(\ref{ab6}) can be replaced by
\begin{equation}\label{ab10}
{\nu_{\eta}}^*=-\frac{\sigma^2\chi\sqrt{m T_0}}{10 n\nu_0 T^{2}(1+c^*(e)/2)}
\int d{\bf v}\int d{\bf v}_1 f^{(0)}({\bf V})f^{(0)}({\bf V}_1)
D_{ij}({\bf V}_1)
\int d\hat{\Vec{\sigma}} (b-1)(V_iV_j+V_{1i}V_{1j})
\end{equation} 
Following the textbook by Chapman and Cowling\cite{chapman},
 the solid angle integral 
over $\hat\sigma$ can be performed. Noting
$(b-1)(V_iV_j+V_{1i}V_{1j})=
\frac{1+e}{2}g_k \hat \sigma_k\{g_l \hat\sigma_l\hat\sigma_i\sigma_j(1+e)
-g_i\hat\sigma_j-g_j\hat\sigma_i\}$ and 
\begin{equation}\label{ab10-1}
D_{ij}({\bf V})\int d\hat{\Vec{\sigma}} (b-1)(V_i V_j+V_{1i}V_{1j})
=\frac{4\pi}{15}(1+e)(e-4)D_{ij}({\bf V}_1)g_ig_j
\end{equation}
we obtain
\begin{equation}\label{ab11}
{\nu_{\eta}}^*=
\frac{2\pi \sigma^2\chi\sqrt{m T_0}}{75 nT^{2}\nu_0}\frac{(1+e)(4-e)}
{(1+c^*(e)/2)}
\int d{\bf V}\int d{\bf V}_1 f^{(0)}({\bf V})f^{(0)}({\bf V}_1)
D_{ij}({\bf V}_1)g_ig_j.
\end{equation}
The result of the integration over ${\bf V}$ and ${\bf V}_1$ leads
to
\begin{equation}\label{ab12}
{\nu_{\eta}}^*=\frac{4\pi n\sigma^2 \chi}{15\nu_0}
\displaystyle\sqrt{\frac{T_0}{m}}(1+e)(4-e).
\end{equation}
In the limit of $e=1$, ${\nu_{\eta}}^*=1$ should be recovered.
Thus, parameters $\chi$ should satisfy
\begin{equation}\label{ab13-2}
\chi=\frac{15 A_m}{8\pi}=0.8176582\cdots,
\end{equation}
which is deviated from 1 a little.
Thus, with the aid of (\ref{nu0})
the final expression becomes eq. (\ref{b40}). This leads to 
the expression of $\eta^*$ as in eq.(\ref{204})

On the other hand, the evaluation of ${\nu_{\kappa}}^*$ and ${\nu_{\mu}}^*$
are as follows.
From the definition of ${\bf S}({\bf V})$ in eq.(\ref{b13})
and the collisional operator $b$ in (\ref{b5}) 
$\int d{\bf V}{\bf S}({\bf V})L[f^{(0)}{\bf S}({\bf V})]$ contains
the solid angle integral
\begin{eqnarray}\label{ab14-1}
& &\int d\hat{\Vec{\sigma}}(b-1)[{\bf S}({\bf V})+{\bf S}({\bf V}_1)]
\nonumber \\
&=&-\frac{m}{2}(1+e)\int d\hat{\Vec{\sigma}}[(1-e)
\frac{({\bf g}\cdot\hat{\Vec{\sigma}})^2}{2}{\bf G}-
(1+e)({\bf g}\cdot\hat{\Vec{\sigma}})^2({\bf G}\cdot\hat{\Vec{\sigma}})\hat{\Vec{\sigma}}
\nonumber \\
& & +({\bf g}\cdot\hat{\Vec{\sigma}})\{({\bf G}\cdot\hat{\Vec{\sigma}}){\bf g}+
({\bf G}\cdot{\bf g})\hat{\Vec{\sigma}}\}]\nonumber \\
&= &\frac{m}{2}(1+e)
[(1+e)\frac{4\pi}{15}(g^2{\bf G}+2({\bf g}\cdot {\bf G}){\bf g})
-\frac{2\pi}{3}(1-e)g^2{\bf G}
-\frac{8\pi}{3}{\bf g}({\bf G}\cdot{\bf g})] \nonumber \\
&=& -\frac{\pi m(1+e)}{15}\{(3-7e)g^2{\bf G}+4(4-e)({\bf g}\cdot{\bf G}){\bf g}\},
\end{eqnarray} 
where ${\bf G}=({\bf V}+{\bf V}_1)/2$.
Thus, the numerator of (\ref{ab7}) becomes
\begin{eqnarray}\label{ab14-2}
\int d{\bf V}{\bf S}({\bf V})\cdot L[f^{(0)}{\bf S}({\bf V})]
&=&\frac{\pi\chi(1+e)}{15}\sigma^2\displaystyle\sqrt{m T_0}
\int d{\bf V}\int d{\bf V}_1f^{(0)}({\bf V})f^{(0)}({\bf V}_1)
\nonumber \\
& &{\bf S}({\bf V}_1)\cdot \{(3-7e)g^2{\bf G}+4(4-e)({\bf g}\cdot{\bf G}){\bf g}\}.
\end{eqnarray}
Here we can rewrite
\begin{equation}\label{ab14-3}
\int d{\bf V}{\bf S}({\bf V})L[f^{(0)}{\bf S}({\bf V})]
=\frac{4\pi \chi \sigma^2n^2}{15m}\displaystyle\sqrt{\frac{T_0}{m}}T^3
(1+e)[\frac{3-7e}{2}I_1+2(4-e)I_2]
\end{equation}
where $I_1$ and $I_2$ are respectively given by
\begin{eqnarray}\label{ab14-4}
I_1&=& \int d{\bf c}\int d{\bf c}_1\tilde f(c)\tilde f(c_1)
({c_1}^2-5/2)({\bf c}\cdot{\bf c}_1+{c_1}^2)|{\bf c}-{\bf c}_1|^2 \nonumber \\
&=& \frac{105}{8} \mu_3-\frac{15}{2}\mu_2+\frac{15}{8}
\end{eqnarray}
and
\begin{eqnarray}\label{ab14-5}
I_2&=& -\int d{\bf c}\int d{\bf c}_1 \tilde f(c)\tilde f(c_1)({c_1}^2-5/2)
(c^2-{c_1^2}){c_1}^2 \nonumber \\
&=& \frac{105}{8}\mu_3-15\mu_2+\frac{45}{8}.
\end{eqnarray}
Substituting (\ref{ab14-4}) and (\ref{ab14-5}) into (\ref{ab14-3}) we
obtain
\begin{equation}
\int d{\bf V}{\bf S}({\bf V})L[f^{(0)}{\bf S}({\bf V})]
=\frac{\pi \chi n^2\sigma^2 T^3}{4m}\displaystyle\sqrt{\frac{T_0}{m}}
(1+e)[7\mu_3(19-11e)-20\mu_2(7-3e)+45-5e].
\end{equation}
Substituting this into (\ref{ab7}) we reach eq.(\ref{b40.1}) .

\section{The distribution function at the first order}

The explicit form of the distribution function at the first order is needed
for the discussion of linear stability of homogeneous cooling states.
Let us assume the form
\begin{equation}\label{c1}
f^{(1)}
=[c_T {\bf S}({\bf V})\cdot\nabla \ln T+c_n {\bf S}({\bf V})\cdot \nabla \ln n
+c_u D_{ij}({\bf V})\nabla_iu_j]f^{(0)}({\bf V}).
\end{equation}
Here recalling eqs.(\ref{b27}),(\ref{b35}) and (\ref{aa11}) we can rewrite
\begin{eqnarray}
\eta &=& -\frac{1}{10}\int d{\bf V}D_{ij}({\bf V}) \tilde C_{ij} ({\bf V})
\nonumber \\
&=& -\frac{c_u}{15}m^2 \int d{\bf V} V^4 f^{(0)} \nonumber \\
&=& -n T^2 c_u \mu_2.
\label{c2}
\end{eqnarray}
From (\ref{aa19}), similarly, we obtain
\begin{eqnarray}
\kappa &=& -\frac{1}{3T}\int d{\bf V} 
{\bf S}({\bf V})\cdot\tilde{\Vec{B}}({\bf V}) \nonumber \\
&=& 
-\frac{c_T}{3T}\int d{\bf V} {\bf S}({\bf V})\cdot {\bf S}({\bf V})f^{(0)}
\nonumber \\
&=& -\frac{5c_T n T^2}{4m}[7\mu_3-10\mu_2+5]
\label{c3}
\end{eqnarray}
and
\begin{eqnarray}
\mu&=& -\frac{1}{3n}\int d{\bf V} 
{\bf S}({\bf V})\cdot\tilde{\Vec{B}}({\bf V}) \nonumber \\
&=& -\frac{c_n}{3n}\int d{\bf V} {\bf S}({\bf V})\cdot {\bf S}({\bf V})f^{(0)}
\nonumber \\
&=& -\frac{5c_n  T^3}{4m}[7\mu_3-10\mu_2+5]
\label{c4}
\end{eqnarray}
Therefore we obtain $f^{(1)}$ as
\begin{equation}\label{c5}
f^{(1)}({\bf V})=-\frac{1}{nT^3}\left[\frac{4m}{5(7\mu_3-10\mu_2+5)}
{\bf S}({\bf V})\cdot(\kappa\nabla T+\mu \nabla n)
+\frac{\eta T}{\mu_2}D_{ij}({\bf V})\nabla_i u_j
\right]f^{(0)}(V).
\end{equation}

\section{Determination of $\zeta^{(2)}$}

In this section, let us determine $\zeta^{(2)}$ which is needed to discuss
the hydrodynamic stability of homogeneous states. The second order correction
$\zeta^{(2)}$ now becomes
\begin{equation}\label{d1}
\zeta^{(2)}=(1-e^2)\frac{2}{3n T}\left[\omega[f^{(1)},f^{(1)}]+
2\omega[f^{(2)},f^{(0)}]\right]
\end{equation}
Now let ${\zeta_L}^{(2)}$ be the linear part of $\zeta^{(2)}$, which satisfies
\begin{equation}\label{d2}
{\zeta_L}^{(2)}=(1-e^2)\frac{4}{3nT}\omega[f^{(0)},f^{(2)}].
\end{equation} 
The second order equation now becomes
\begin{equation}\label{d3}
({\partial_t}^{(0)}+L)f^{(2)}=-{\partial_t}^{(2)}f^{(0)}
-({D_t}^{(1)}+{\bf V}\cdot \nabla)f^{(1)}+J[f^{(1)},f^{(1)}].
\end{equation}
Here the contribution from $J[f^{(1)},f^{(1)}]$ can be neglected, because
this term creates nonlinear terms of hydrodynamic variables.
We also note 
\begin{eqnarray}
{\partial_t}^{(2)}n &=& 0 \nonumber \\
m n {\partial_t}^{(2)}{\bf u} &=& \frac{\eta}{3}\nabla(\nabla\cdot {\bf u})
+\eta \nabla^2{\bf u} \nonumber \\
{\partial_t}^{(2)}T &=& -T \zeta^{(2)}+\frac{2}{3n}
(\kappa \nabla^2T+\mu \nabla^2 n)
\label{d3-1}
\end{eqnarray}
for linearized equations.
Taking into account the above argument\cite{chapman} and (\ref{b24}) we obtain
\begin{eqnarray}\label{d4}
({\partial_t}^{(0)}&+&L){f_L}^{(2)}-{\zeta_L}^{(2)}T \partial_T f^{(0)}
\nonumber \\
&=&-\frac{2}{3n}(\kappa\nabla^2 T+\mu\nabla^2n)\partial_Tf^{(0)} \nonumber \\
& & +\frac{4m {\bf S}({\bf V}){\bf V}:
(\kappa\nabla\nabla T+\mu \nabla\nabla n )}{5nT^3(7\mu_3-10\mu_2+5)}f^{(0)}
\nonumber \\
& & +\frac{\eta}{nm}[\frac{1}{3}\nabla (\nabla\cdot{\bf u})+\nabla^2{\bf u}]
\cdot\frac{\partial}{\partial {\bf V}}f^{(0)} \nonumber \\
& & -\frac{4m}{5 n T^3(7\mu_3-10\mu_2+5)}\left(\frac{2}{3}T\kappa+n \mu\right)
{\bf S}({\bf V})\cdot(\nabla\nabla\cdot {\bf u}) \nonumber \\
& & -\frac{\eta}{n T^2\mu_2}D_{ij}({\bf V})[(nm)^{-1}\nabla_i\nabla_j p
-{\bf V}:\nabla(\nabla_iu_j)]f^{(0)}.
\end{eqnarray}
Further simplification is possible by noting that $\zeta^{(2)}$ is a scalar
so that any contributions to ${f_L}^{(2)}$ that are vector or traceless 
functions does not contribute. Let $\delta {f_L}^{(2)}$ be the residual
part of ${f_L}^{(2)}$.
Thus, we can simplify
\begin{eqnarray}
\label{d5}
({\partial_t}^{(0)}&+&L)\delta {f_L}^{(2)}-{\zeta_L}^{(2)}\partial_Tf^{(0)}
\nonumber \\
&=&[\frac{4m}{15n T^3}\frac{f^{(0)}({\bf V}){\bf S}({\bf V})\cdot {\bf V}}
{(7\mu_3-10\mu_2+5)}-\frac{2}{3nT}T\partial_Tf^{(0)}]
(\kappa\nabla^2T+\mu \nabla^2n).
\end{eqnarray}
Let us assume
\begin{equation}\label{d6}
\delta{f_L}^{(2)}=M(T,{\bf V})\nabla^2T +N(T,{\bf V})\nabla^2n
\end{equation}
and taking into account (\ref{b43}) we obtain
\begin{equation}\label{d8}
{\partial_t}^{(0)}\nabla^2 T=-\zeta^{(0)}\nabla^2 T-
\frac{T \zeta^{(0)}}{n}\nabla^2n.
\end{equation}
Substituting this into (\ref{d6}) eq.(\ref{d5}) becomes
\begin{equation}\label{d9}
(-\zeta^{(0)}T\partial_T-\zeta^{(0)}+L)M-\zeta_1T \partial_Tf^{(0)} 
=\kappa\left[\frac{4m}{15nT^3}
\frac{f^{(0)}{\bf S}({\bf V})\cdot {\bf V}}{7\mu_3-10\mu_2+5}
-\frac{2}{3nT}T \partial_Tf^{(0)}\right]
\end{equation}
and
\begin{equation}\label{d10}
(-\zeta^{(0)}T\partial_T+L)M-\zeta_2T \partial_Tf^{(0)} 
= \frac{T \zeta^{(0)}}{n}M+
\mu\left[\frac{4m}{15nT^3}
\frac{f^{(0)}{\bf S}({\bf V})\cdot {\bf V}}{7\mu_3-10\mu_2+5}
-\frac{2}{3nT}T \partial_Tf^{(0)}\right].
\end{equation}
The scalar functions $M$ and $N$ are orthogonal to 
1, ${\bf v}$ and ${\bf V}^2$ 
from the solvability conditions (\ref{sol1})-(\ref{sol3})
, {\it i.e.} not to contribute the transport
coefficients determined in Appendix C. Thus, we can assume
\begin{equation}\label{d11}
M={c_T}^{(2)}Q(c)f^{(0)}({\bf V}), \quad
N={c_n}^{(2)}Q(c)f^{(0)}({\bf V})
\end{equation}  
with the form
\begin{equation}\label{d12}
Q(c)=c^4+\alpha c^2+\beta. 
\end{equation}
From the orthogonal conditions, $\alpha$ and $\beta$ can be determined as
(\ref{b43-2}).
Thus, we obtain (\ref{b43-1}).

Equations to derive ${c_T}^{(2)}$ and ${c_n}^{(2)}$ are obtained from
eqs.(\ref{d9}) and (\ref{d10}). At first they are rewritten as
\begin{eqnarray}
\frac{\zeta^*}{2}\frac{d}{d{\bf c}}&\cdot&({c_T}^{(2)}{\bf c}Q\tilde f)
-\zeta^*{c_T}^{(2)}Q\tilde f+\frac{{c_T}^{(2)}}{\nu_0}L[Q\tilde f]
+\frac{{\zeta_1}^*}{2\nu_0}\frac{d}{d{\bf c}}\cdot ({\bf c}\tilde f)
\nonumber \\
&=&\frac{\kappa}{3p \nu_0}\left[
\frac{8\tilde f c^2(c^2-5/2)}{5(7\mu_3-10\mu_2+5)}
+\frac{d}{d{\bf c}}\cdot ({\bf c}\tilde f)
\right]
\end{eqnarray}
and
\begin{eqnarray}
\frac{\zeta^*}{2}\frac{d}{d{\bf c}}&\cdot&({c_n}^{(2)}{\bf c}Q\tilde f)
+\frac{{c_n}^{(2)}}{\nu_0}L[Q\tilde f]
+\frac{{\zeta_2}^*}{2\nu_0}\frac{d}{d{\bf c}}\cdot ({\bf c}\tilde f)
\nonumber \\
&=&\frac{\zeta^*T}{n}{c_T}^{(2)}Q\tilde f+
\frac{\mu}{3nT\nu_0}\left[
\frac{8\tilde f c^2(c^2-5/2))}{5(7\mu_3-10\mu_2+5)}
+\frac{d}{d{\bf c}}\cdot ({\bf c}\tilde f)
\right].
\end{eqnarray}
Multiplying them  by $c^4$ and
integrate them all over ${\bf c}$ we obtain
\begin{eqnarray}
(-3\zeta^*+{\nu_{\zeta}}^*){c_T}^{(2)}
&=&\frac{16}{\nu_0}\left(\frac{\zeta_1}{2}-\frac{\kappa}{3p}\right)
\frac{\mu_2}{63\mu_4+14\alpha \mu_3+4\beta \mu_2} \nonumber \\
& &+\frac{8\kappa}{15p\nu_0(7\mu_3-10\mu_2+5)}
\frac{63\mu_4-35\mu_3}{63\mu_4+14\alpha \mu_3+4\beta \mu_2}
\label{d15}
\end{eqnarray}
\begin{eqnarray}
(-2\zeta^*+{\nu_{\zeta}}^*){c_n}^{(2)}
&=&\frac{16}{\nu_0}\left(\frac{\zeta_2}{2}-\frac{\mu}{3p}\right)
\frac{\mu_2}{63\mu_4+14\alpha \mu_3+4\beta \mu_2} \nonumber \\
& &+\frac{8\mu}{15p\nu_0(7\mu_3-10\mu_2+5)}
\frac{63\mu_4-35\mu_3}{63\mu_4+14\alpha \mu_3+4\beta \mu_2},
\label{d16}
\end{eqnarray}
where 
\begin{eqnarray}
{\nu_{\zeta}}^*&\equiv& \frac{\int d{\bf c}c^4 L[Q(c)\tilde f(c)]}
{\nu_0 \int d{\bf c} c^4 Q(c)\tilde f(c)} \nonumber \\
&=& \frac{4W
}{15 (\frac{63}{4}\mu_4+\frac{7}{2}\alpha \mu_3+\beta \mu_2)}.
\end{eqnarray}
Here $W$ is defined as $W\equiv \int d{\bf c} c^4 L[Q(c)\tilde f(c)]/\nu_0$.
Introducing dimensionless ${c_T}^*$ and ${c_n}^*$ by
\begin{equation}
{c_T}^*\equiv \frac{{c_T}^{(2)}}{\kappa_0/p\nu_0},\quad
{c_n}^*\equiv \frac{{c_n}^{(2)}}{\kappa_0/n^2\nu_0}
\end{equation}
eqs.(\ref{d15}) and (\ref{d16}) can be rewritten as
\begin{eqnarray}\label{d15a}
{c_T}^*&=&\frac{1}{({\nu_{\zeta}}^*-3\zeta^*)
(63\mu_4+14\alpha\mu_3+4\beta\mu_2)} \nonumber \\
& &\times \left[\frac{8(63\mu_4-35\mu_3)\kappa^*}{15(7\mu_3-10\mu_2+5)}
+\frac{16\mu_2}{3}({\zeta_1}^*-\kappa^*)\right]
\end{eqnarray}
and
\begin{eqnarray}\label{d16a}
{c_n}^*&=&\frac{1}{({\nu_{\zeta}}^*-2\zeta^*)
(63\mu_4+14\alpha\mu_3+4\beta\mu_2)} \nonumber \\
& &\times \left[\frac{8(63\mu_4-35\mu_3)\mu^*}{15(7\mu_3-10\mu_2+5)}
+\frac{16\mu_2}{3}({\zeta_2}^*-\kappa^*)\right].
\end{eqnarray}
Eliminating ${\zeta_1}^*$ and ${\zeta_2}^*$ from (\ref{d15a}) and
(\ref{d16a}) we obtain (\ref{zeta1,2}).

Now, we can evaluate  $W$ as in the previous
section. The result may be
\begin{equation}\label{w}
W=
\frac{(1-e^2)}{8}\left[\left(\frac{37}{6}+\frac{e^2}{2}\right)J_1-
3(1-e^2)
\right],
\end{equation}
where
\begin{equation}\label{j1}
J_1=\frac{15}{4}\left[\mu_2(\frac{15}{4}\mu_2+\frac{3}{2}\alpha+2\beta)+
\frac{63}{4}\mu_4+\frac{7}{2}\alpha\mu_3\right]
\end{equation}
and
\begin{equation}\label{j2}
J_2=\frac{3}{2}\left(\frac{105}{8}\mu_3+\frac{15}{4}\alpha \mu_2+\frac{3}{2}\beta\right).
\end{equation}
Thus ${\nu_{\zeta}}^*$ is determined by
\begin{equation}\label{e22}
{\nu_{\zeta}}^*
=\frac{4}{15}
\frac{W}{\frac{63}{4}\mu_4+\frac{7}{2}\alpha\mu_3+\beta \mu_2}.
\end{equation}

\end{document}